\documentclass[12pt,reqno]{amsart}
\usepackage{latexsym}

\renewcommand{\Re}{{\operatorname{Re}}}

\renewcommand{\epsilon}{\varepsilon}
\renewcommand{\Re}{{\operatorname{Re}}}

\newcommand{\sym}{{\operatorname{Sym}}}

\newcommand{\crit}{{\operatorname {crit}}}

\newcommand{\sm}{\setminus}
\newcommand{\szego}{Szeg\"o }

\newcommand{\inv}{^{-1}}
\newcommand{\kahler}{K\"ahler }

\newcommand{\wt}{\widetilde}
\newcommand{\wh}{\widehat}

\newcommand{\R}{{\mathbb R}}

\newcommand{\C}{{\mathbb C}}

\newcommand{\Z}{{\mathbb Z}}

\renewcommand{\d}{\partial}

\newcommand{\U}{{\rm U}}

\renewcommand{\phi}{\varphi}

\newcommand{\ccal}{\mathcal{C}}
\newcommand{\dcal}{\mathcal{D}}
\newcommand{\ecal}{\mathcal{E}}

\newcommand{\hcal}{\mathcal{H}}
\newcommand{\ical}{\mathcal{I}}
\newcommand{\kcal}{\mathcal{K}}

\newcommand{\qcal}{{\bf Q}}

\newcommand{\mcal}{\mathcal{M}}

\newcommand{\fcal}{\mathcal{F}}
\newcommand{\lcal}{\mathcal{L}}
\newcommand{\ncal}{\mathcal{N}}
\newcommand{\pcal}{\mathcal{P}}

\newcommand{\ocal}{\mathcal{O}}
\newcommand{\scal}{\mathcal{S}}

\newcommand{\La}{\Lambda}
\newcommand{\la}{\lambda}
\newcommand{\ep}{\varepsilon}
\newcommand{\de}{\delta}

\newtheorem{theo}{{\sc Theorem}}[section]

\newtheorem{prop}[theo]{{\sc Proposition}}

\title[Counting string/M vacua]
{ Counting string/M vacua}

\author{Steve Zelditch}
\address{Department of Mathematics, Johns Hopkins University, Baltimore,
MD 21218, USA} \email{zelditch@math.jhu.edu}

\thanks{Research partially supported by
NSF grant DMS-0302518}

\date{March 3, 2005}

\begin{document}

\begin{abstract} We report on some recent work with M. R. Douglas and B. Shiffman on vacuum statistics
for flux compactifications in string/M theory.

\end{abstract}

\maketitle

\tableofcontents

\section{Introduction} According to string/M  theory, the vacuum state of our universe is a  $10$ dimensional
spacetime of the form  $M^{3,1} \times
X$, where $M^{3,1}$ is Minkowski space and $X$ is a
small $3$-complex dimensional Calabi-Yau manifold $X$ known as the
`small' or `extra' dimensions \cite{CHSW, St}. The {\it vacuum selection problem} is that there
are many candidate vacua for the  Calabi-Yau $3$-fold $X$. Here, we report on recent joint work with B. Shiffman and M. R. Douglas
devoted to  counting the number of supersymmetric vacua of type IIb flux compactifications \cite{DSZ, DSZ2, DSZ3}. We also describe closely related
the physics articles of Ashok-Douglas and Denef-Douglas \cite{D, AD, DD} on the same problem.

At the time of writing of this article, vacuum statistics is being intensively investigated by many string theorists (see
for instance \cite{DGKT, CQ, GKT, S, Ar} in addition to the articles cited above). One often hears that the number of possible
vacua is of order $10^{500}$ (see e.g. \cite{BP}). This large figure is sometimes decried (at this time) as a blow to predictivity
of string/M theory or extolled as giving string theory a  rich enough `landscape' to contain vacua that match the physical parameters (e.g. the
cosmological constant) of our universe. However, it is very difficult to obtain sufficiently  accurate results on vacuum counting
 to  justify  the claims of $10^{500}$ total vacua, or
even the existence of one vacuum which is consistent with known physical parameters. The purpose of our work is to
develop methods and results relevant to
 accurate vacuum counting.

From a mathematical viewpoint, supersymmetric vacua are critical points
\begin{equation} \label{CPEQ} \nabla W_G(Z) = 0 \end{equation} of certain holomorphic sections $W_G$ called {\it flux superpotentials} of
a line bundle $\lcal \to \ccal$ over the moduli space $\ccal$ of complex structures on $X \times T^2$ where $T^2 = \R^2/\Z^2$. Flux superpotentials
depend on a choice of flux $G \in H^3(X, \Z \oplus \sqrt{-1} \Z)$. There is a constraint on $G$ called the `tadpole constraint',
so that $G$ is a lattice point lying in  a certain  hyperbolic shell $0 \leq Q[G] \leq L$ in $H^3(X, \C)$ (\ref{HR}). Our goal is to count all critical
points of all flux superpotentials $W_G$ in a given compact set of $\ccal$  as $G$ ranges over such lattice points.
Thus, counting vacua in $K \subset \ccal$  is a combination of an  equidistribution problem for projections of lattice points and
an equidistribution problem  for
critical points of  random holomorphic sections.

 The work we report on gives a rigorous foundation for the
program initiated by M. R. Douglas  \cite{D} to count vacua by making an approximation to the Gaussian ensembles the other two authors
 were using to study statistics of  zeros of random holomorphic sections  (cf. \cite{SZ, BSZ}).
  The results we describe here are the first rigorous results on counting vacua in a reasonably general
 class of models (type IIb flux compaticifications). They are admittedly still in a rudimentary stage, in particular because
 they are asymptotic rather than effective. We will discuss the difficulties in making them effective below.

This report is a written version of our talk at the QMath9 conference
in Giens in October, 2004. A more detailed expository article with background on statistical algebraic
geometry as well as string theory is given in  \cite{Z}, which was
 based on the author's AMS address in Atlanta, January 2005.

\section{Type IIb flux compactifications of string/M theory }

The string/M theories we consider are type IIb string theories  compactified on a complex $3$-dimensional Calabi-Yau manifold $X$ with flux
\cite{GKP, GVW,GKTT, GKT, AD}.
 We recall that a Calabi-Yau $3$-fold is
a compact complex manifold $X$ of dimension $3$
with trivial canonical bundle $K_X$, i.e. $c_1(X) = 0$ \cite{Gr, GHJ}. Such $X$ possesses  a unique Ricci flat \kahler
metric in each \kahler class. In what follows, we fix the \kahler class, and then the  CY metrics correspond to the complex structures
on $X$.
  We denote the moduli space of complex structures on $X$ by $\mcal_{\C}$. In addition to the complex structure moduli on $X$ there is an extra parameter $\tau$ called the dilaton
axion, which ranges over complex structure moduli on $T^2 = \R^2/\Z^2$. Hence, the full  configuration space $\ccal$ of the model is the
product
\begin{equation} \label{CCAL} \ccal = \mcal_{\C} \times \ecal,\;\;( Z= (z, \tau);\;\; z \in \mcal_{\C}, \tau \in \ecal) \end{equation}
where $\ecal = \hcal/ SL(2, \Z)$ is the moduli space of complex
 $1$-tori (elliptic curves).  One can think of
 $\ccal$ as a moduli space of complex structures on the CY $4$-fold $X \times T^2$.

By   `flux' is meant a complex integral $3$-form
\begin{equation} \label{G} G = F + i H  \in H^3(X, \Z \oplus \sqrt{-1} \Z). \end{equation}
The   {\it flux superpotential} $W_G(Z)$ corresponding to $G$ is defined as follows:
On a Calabi-Yau $3$-fold, the space $H^{3,0}_z(X)$ of holomorphic $(3,0)$-forms for each complex structure $z$
on $X$ has dimension $1$, and
we denote a holomorphically varying family by  $\Omega_z \in H^{3,0}_z(X)$.
 Given $G$ as in (\ref{G}) and $  \tau \in \hcal$, physicists
 define the superpotential corresponding to $G, \tau$    by:
\begin{equation} \label{WSUBG}  W_{G}(z, \tau) = \int_X ( F - \tau H)  \wedge \Omega_{z}. \end{equation}
This is not  well-defined as a function on $\ccal$, since $\Omega_z$ is not unique and $\tau$ corresponds to the
holomorphically varying form $\omega_{\tau} = dx + \tau dy \in H^{1,0}_{\tau}(T^2)$ which is not unique either.
To be more precise,  we define $W_G$ to be  a holomorphic
 section of a line bundle $\lcal \to \ccal, $ namely the dual line bundle to the Hodge line bundle  $H^{4,0}_{z, \tau} = H^{3,0}_z (X)
 \otimes H^{1,0}(T^2) \to \ccal$.
We form the $4$-form on $X \times T^2$
$$\tilde{G} =  F \wedge dy + H \wedge dx$$ and define a linear functional on $H^{3, 0}_z(X) \otimes H^{1,0}_{\tau}(T^2)$
by
\begin{equation} \label{WG} \langle W_G(z, \tau), \Omega_z \wedge \omega_{\tau} \rangle = \int_{X \times T^2} \tilde{G} \wedge  \Omega_z \wedge \omega_{\tau}.
\end{equation}
When $\omega_{\tau} = dx + \tau dy$ we obtain the original formula.
As $Z = (z, \tau)\in \ccal$ varies, (\ref{WG}) defines  a holomorphic section of the line
bundle $\lcal $ dual to $H^{3,0}_z \otimes H^{1, 0}_{\tau}\to \ccal$.

The Hodge bundle carries a
natural Hermitian metric
$$h_{WP}(\Omega_{z} \wedge \omega_{\tau}, \Omega_{z} \wedge \omega_{\tau}) = \int_{X \times T^2}
\Omega_{z} \wedge \omega_{\tau} \wedge \overline{\Omega_{z} \wedge \omega_{\tau}} $$
known as the Weil-Petersson metric, and an associated metric  (Chern)
 connection by $\nabla_{WP}.$ The \kahler potential of the Weil-Petersson metric on $\mcal_{\C}$ is defined by
 \begin{equation} \label{KP} \kcal = - \ln \langle \Omega, \Omega \rangle = \ln \left(-i
\int_X \Omega \wedge \overline{\Omega} \right).\end{equation}
There is a similar definition on $\ecal$ and we take the direct sum to obtain a \kahler metric on $\ccal.$
We  endow $\lcal$
with the dual Weil-Petersson metric and connection.
The hermitian line bundle  $(H^{4,0}, h_{WP}) \to \mcal_{\C}$ is a positive line
bundle, and it follows that $\lcal$ is a negative line bundle.

The vacua we wish to count are the classical vacua of the effective supergravity Lagrangian of the string/M model, which is derived by `integrating
out' the massive modes (cf. \cite{St}). The only term relevant of the Lagrangian to our counting problem is the scalar potential \cite{WB}
\begin{equation}\label{VG}  V_G(Z) = |\nabla W_G(Z)|^2 - 3 |W(Z)|^2, \end{equation}
where the connection and hermitian metric are the Weil-Petersson ones.
We only consider the supersymmetric vacua here, which are the special critical points $Z$ of $V_G$ satisfying
(\ref{CPEQ}).

\section{Critical points and Hessians of holomorphic sections}

We see that type IIb flux compactifications involve holomorphic sections of hermitian holomorphic line bundles
over complex manifolds. Thus, counting flux vacua is a problem in complex geometry. In this section, we provide
a short review  from \cite{DSZ, DSZ2}.

Let $L \to M$ denote a holomorphic line bundle over a complex manifold, and endow $L$ with
a hermitian metric $h$. In a local frame $e_{L}$ over an open set $U \subset M$, one
defines the \kahler potential $K$ of $h$  by
\begin{equation}\label{Kpot}|e_L(Z)|_h^2 =
e^{-K(Z)}  \;. \end{equation} We
write a section $s \in H^0(M,L) $ locally as $s=fe_{L}$ with $f \in \ocal(U)$. We further
choose local coordinates $z$. In this frame and local coordinates,
the covariant derivative of a section $s$ takes the local form
\begin{equation}\label{covar}
\nabla s=\sum_{j=1}^m\left( \frac{\d f}{\d Z_j} -f \frac{\d K}{\d
Z_j}\right)dZ_j \otimes e_{\lcal}=\sum_{j=1}^m e^{K}\frac{\d }{\d
Z_j}\left(e^{-K}\, f\right)dZ_j \otimes e_{\lcal}\;.
\end{equation}
The critical point equation $\nabla s(Z) = 0$  thus reads,
$$ \frac{\d f}{\d Z_j} -f \frac{\d K}{\d
Z_j} = 0. $$

It is important to observe that although $s$ is holomorphic, $\nabla s$ is not, and the critical point
equation is only $C^{\infty}$ and not holomorphic. This is due to the factor $\frac{\d K}{\d Z_j}$, which
is only smooth. Connection critical points of $s$ are the same as ordinary critical
points of $\log |s(Z)|_{h}$. Thus, the critical point equation is a system of real equations and the number of critical
points varies with the holomorphic section. It is not a topological invariant, as would be the number
of zeros  of independent $m$ sections in dimension $m$,  even on a compact complex manifold.
This is one reason why counting critical points, hence vacua, is so complicated.

We now consider the Hessian of a section at a critical point.
The Hessian of a holomorphic section $s$ of a general Hermitian
holomorphic  line bundle $(L, h) \to M$  at a critical point $Z$
is the tensor
$$D \nabla W(Z) \in T^* \otimes T^* \otimes L$$
where $D$ is a connection on $T^* \otimes L$. At a critical
point $Z$, $D \nabla s(Z)$ is independent of the choice of
connection on $T^*$.
The Hessian $D \nabla W(Z)$ at a critical point determines the
complex symmetric matrix $H^c$ (which we call the `complex
Hessian'). In an adapted local frame (i.e. holomorphic derivatives vanish at $Z_0$) and in \kahler normal coordinates, it takes the form
\begin{equation}\label{HmatriX} H^c:=
\begin{pmatrix} H' &H''\\[6pt] \overline{H''} &\overline{H'}
\end{pmatrix} =\begin{pmatrix} H' &-f(Z_0)\Theta\\[8pt]
-\overline{f(z_0)\Theta} &\overline{H'}
\end{pmatrix}\;,
\end{equation}  whose components  are given by
\begin{eqnarray}H'_{jq} &=& (\frac{\d}{\d
Z_j} - \frac{\d K}{\d Z_j}) (\frac{\d}{\d Z_q} - \frac{\d K}{\d Z_q}) f(Z_0)\;,\label{H'}\\
    H''_{jq} &=& -\left.f \frac{\d^2 K}{\d Z_j\d\bar
Z_q}\right|_{Z_0}=-f(Z_0)\Theta_{jq}\,.\label{H''}\end{eqnarray}
Here, $
\Theta_h(z_0)=\sum_{j,q}\Theta_{jq}dZ_j\wedge d\bar
Z_q$ is the curvature form.

\section{The critical point problem}

We can now define  the critical point equation (\ref{CPEQ}) precisely.  We define a
 supersymmetric  vacuum of the flux superpotential $W_G$ corresponding to the flux
 $G$ of (\ref{G}) to be a critical
point $\nabla_{WP} W_G(Z) = 0$ of $W_G$ relative to the Weil-Petersson connection on $\lcal$.

We obtain a local formula by writing $W_G(Z) = f_G (Z) e_Z$ where $e_Z$ is local frame
for $\lcal \to \ccal$. We choose the local frame $e_Z$ to be dual to  $\Omega_z \otimes \omega_{\tau}$,
and then $f_G(z, \tau)$ is given by the formula (\ref{WSUBG}). The $\ecal$ component of $\nabla_{WP}$
is $\frac{\partial}{\partial \tau} - \frac{1}{\tau - \bar{\tau}}$.  The critical point equation is  the system:
\begin{equation} \label{CPE} \left\{ \begin{array}{l} \int_X (F - \tau H) \wedge
\{\frac{\partial \Omega_z}{\partial z_j} +  \frac{\partial K}{\partial z_j} \Omega_z\}= 0,\\ \\
\int_X (F - \bar{\tau} H) \wedge \Omega_z = 0, \end{array} \right.
 \end{equation} where $K$ is from (\ref{KP}).

Using the {\it special geometry} of $\ccal$ (\cite{St3, Can1}), one finds  that the critical point equation
is equivalent to the following restriction on the Hodge decomposition of $H^3(X, \C)$ at $z$:
\begin{equation} \label{CPEH} \nabla_{WP} W_G (z, \tau) = 0 \iff
F - \tau H \in H^{2, 1}_z \oplus H_z^{0, 3}. \end{equation}
Here, we recall that each complex structure  $z \in \mcal_{\C}$ gives rise to  a Hodge decomposition
\begin{equation} \label{HD} H^3(X, \C) = H^{3,0}_z(X) \oplus H^{2,1}_z (X)\oplus H^{1,2}_z(X) \oplus H^{0, 3}_z(X) \end{equation}
into forms of type $(p,q)$. In the case of a CY $3$-fold,
$h^{3,0} = h^{0, 3} = 1$,  $h^{1,2} = h^{2,1}$ and $b_3 = 2 + 2 h^{2,1}.$

Next, we specify the tadpole constraint.
We define the  real symmetric  bilinear form on $H^3(X, \C)$ by
\begin{equation} \label{HR} Q(\psi, \phi) = i^3  \int_X \psi \wedge \bar{\phi}. \end{equation}  The Hodge-Riemann bilinear relations
for  a $3$-fold say that the form  $Q$ is definite in each
$H^{p, q}_z(X)$ for $p + q = 3$ with sign alternating $ + - + -$ as one moves left to right in (\ref{HD}).
The tadpole constraint is that
\begin{equation} \label{TC} Q[G] = i^3  \int_X G \wedge \bar{G}\leq L. \end{equation}
Here, $L$ is determined by $X$ in a complicated way (it equals $\chi(Z)/24$ where $Z$ is CY $4$-fold which is  an elliptic fibration over $X/g$,
where $\chi(Z)$ is the Euler characteristic and where $g$ is an involution of $X$).
Although $Q$ is an indefinite symmetric bilinear form,
we see  that $Q >> 0$ on $H^{2,1}_z(X) \oplus H^{0, 3}_z$ for any complex structure $z$.

We now explain the sense in which we are dealing with a lattice point problem. The definition of
 $W_G$  makes sense for
any $G \in H^3(X, \C)$, so we obtain a real (but not complex) linear  embedding  $H^3(X, \C)  \subset H^0(\ccal,
\lcal)$. Let us   denote the image by $\fcal$ and call it  the space of   complex-valued  flux
superpotentials with dilaton-axion.   The set of $W_G$  with $G \in H^3(X, \Z \oplus \sqrt{-1} \Z)$
is then a  lattice $\fcal_{\Z} \subset \fcal$, which we will call  the lattice of quantized (or integral) flux superpotentials.

Each integral flux  superpotential $W_G$ thus gives rise to a discrete set  of critical points  $Crit(W_G) \subset \ccal$, any of which could be the vacuum
state of the universe.
Moreover, the flux $G$ can be any element of $H^3(X, \Z \oplus \sqrt{-1} \Z)$ satisfying the  tadpole constraint (\ref{TC}). Thus,
the set of possible vacua is  the union  \begin{equation}\label{VACUA}
 \mbox{Vacua}_L\; = \bigcup_{G \in H^3(X, \Z \oplus \sqrt{-1} Z),\;\; 0 \leq Q[G] \leq L} \;
\mbox{Crit}(W_G). \end{equation}
Our purpose is to count the number of vacua $\# \mbox{Vacua}_L \cap K$ in any given compact subset $K \subset \ccal.$

 More generally, we wish to consider the sums
 \begin{equation} \label{NPSI}  N_{\psi}(L) =  \sum_{N \in H^3(X, \Z \oplus
\sqrt{-1} \Z) : Q[N] \leq L} \langle C_N, \psi \rangle, \end{equation} where
\begin{equation} \label{CN} \langle C_N, \psi \rangle = \sum_{(z, \tau): \nabla N(z, \tau) = 0} \;
\psi(N, z, \tau),  \end{equation}  and where $\psi$ is a reasonable function on the
 incidence relation
\begin{equation} \label{ANOTHERICAL} \ical = \{(W; z, \tau) \in \fcal \times \ccal: \nabla W(z, \tau) =
0\}. \end{equation}
We often write $Z = (z, \tau) \in \ccal$. Points $(W, Z)$ such that $Z$ is a degenerate critical point of $W$ cause
problems.
They belong to the {\it discriminant variety}  $\wt \dcal\subset \ical$ of  singular points of the projection
$\pi: \ical \to \fcal$. We note that $\pi^{-1}(W) = \{(W, Z): Z \in Crit(W)\}$. This number is constant
on each component of $ \fcal\sm \dcal$ where
$\dcal = \pi(\wt\dcal)$ but jumps as we cross over $\dcal$.

To  count critical points in
a compact subset $\kcal
\subset \ccal$  of moduli space,  we would   put $\psi = \chi_K(z, \tau)$. We often want
to exclude degenerate critical points and then use test functions $\psi(W, Z)$ which are homogeneous
of degree $0$ in $W$ and vanish on $\wt \dcal$  Another
important example is the cosmological constant $\psi(W, z, \tau) = V_W(z, \tau),$ i.e.
 the value of the potential at the vacuum, which is homogeneous of degree $2$ in $W$.

 \section{Statement of results}

 We first state an initial estimate which is regarded as `trivial'
in lattice counting problems. In pure lattice point problems it is sharp, but we  doubt that it is sharp
in the vacuum counting problem because of the `tilting' of the projection  $\ical \to \ccal$.
We  denote by $\chi_Q$ the characteristic function of the hyperbolic shell  $0 < Q_Z[W] < 1 \subset \fcal$
and by  $\chi_{Q_Z}$ the characteristic function of the elliptic
shell $0 < Q_Z[W] < 1 \subset \fcal_Z$.

\begin{prop} \label{LNMO} Suppose that $\psi(W, Z) = \chi_K$ where $K \subset \ical$ is an
open set with smooth boundary. Then:
$$\ncal_{\psi}(L) = L^{b_3}\left[
 \int_{\ccal}
\int_{\fcal_{Z}} \psi(W, Z)  |\det H^c W(Z)|
  \chi_{Q_{Z}} dW dV_{WP}(Z)+  R_{K}(L) \right],
$$
where $dV_{WP}$ is the Weil-Petersson volume form and where:

\begin{enumerate}

\item If $\overline K$ is disjoint from the $\wt \dcal$,
then $R_K(L) =   O\left(L^{-1/2}\right).  $

\item If $\overline K$ is a general compact set (possibly
intersecting  the discriminant locus), then $R_K(L) =
O\left(L^{-1/2}\right) $

\end{enumerate}

\end{prop}

Here, $b_3 =  \dim H_3(X, \R), Q_Z = Q_{z, \tau} = Q|_{\fcal_{z, \tau}}$, and $\chi_{Q_{Z}}(W) $ is the characteristic
function of $\{Q_{z, \tau} \leq 1\} \subset \fcal_{z, \tau}. $ Also,
 $H^c  W(Z)$ is the complex Hessian of $W$ at the critical point
$Z$ in the sense of  (\ref{HmatriX}).   We note
that the integral converges since $\{Q_Z \leq 1\}$ is an ellipsoid
of finite volume. This is an asymptotic formula which is a good estimate on the number of vacua  when
$L$ is large (recall that $L$ is a topological invariant determined by $X$).

The reason for assumption (1)  is that number of
critical points and the summand  $\langle C_W, \psi \rangle$ jump across
$\dcal$, so in $N_{\psi}(L)$ we are  summing a discontinuous function. This discontinuity could cause a relatively large
error term in the asymptotic counting.
However,  superpotentials of physical
interest have non-degenerate supersymmetric critical points. Their Hessians at the critical points are
`fermionic mass matrices', which in physics have only  non-zero eigenvalues (masses), so it is reasonable
 assume that $supp \psi$ is disjoint
from $\dcal.$

Now we state the main result.

\begin{theo} \label{STRINGMAIN} Suppose $\psi(W, z, \tau) \in C^{\infty}_b(\fcal \times \ccal)$ is homogeneous
of degree $0$ in $W$ , with  $\psi(W, z, \tau) = 0$ for $W \in \dcal$. Then
$$\ncal_{\psi}(L) = L^{b_3}\left[\int_{\ccal}  \int_{\fcal_{z, \tau}} \psi(W, z, \tau) |\det H^c W(z, \tau)|
  \chi_{Q_{z, \tau}}(W)  dW dV_{WP}(z, \tau)
   + O\left(L^{-\frac{2 b_3}{2b_3+1}}\right)\right].
$$
Here, $C^{\infty}_b$ denotes bounded smooth functions.
\end{theo}

There is a simple generalization to homogeneous functions of any degree such as the cosmological constant.
The formula is only the starting point of a number of further versions which will be presented in \S \ref{FURTHER}
 in which we `push-forward' the $dW$ integral
under the Hessian map, and then perform an Itzykson-Zuber-Harish-Chandra transformation on the integral. The latter
version gets rid of the absolute value and seems to most useful for numerical studies. Further, one can use the
special geometry of moduli space to simplify the resulting integral. Before discussing them, we pause to compare
our results to the expectations in the string theory literature.

\section{Comparison to the physics literature}

The reader following the developments in string theory may have encountered
discussions of the `string theory landscape' (see e.g. \cite{S,  BP}).
The  multitude of  superpotentials and vacua is  a problem for the predictivity of  string theory. It is possible
that a unique vacuum will distinguish itself in the future, but until then  all critical points are candidates for the small dimensions
of the universe, and several groups of physicists are counting or enumerating them in various models (see e.g. \cite{DD, CQ, DGKT}).

The graph of the scalar potential energy  may be visualized  as a landscape \cite{S} whose local
minima are the possible  vacua. It is common to hear that there are roughly  $10^{500}$ possible vacua.
This  heuristic figure appears to originate in the following reasoning:
assuming $b_3 \sim 250$, the potential energy $V_G(Z)$ is a function
roughly $500$ variables (including fluxes $G$).
 The critical point equation for a function of $m$ variables is a system of
 $m$ equations. Naively,  the number of solutions should grow like $d^m$ where $d$ is the number of solutions of
the $j$th equation with the other variables held fixed. This would follow from B\'ezout's formula if the function was
a polynomial and if we were counting complex zeros. Thus, if the `degree' of $V_G$ were a modest figure of $10$ we would
obtain the heuristic figure.

  Such an exponential growth rate of critical
points in the number of variables  also arises
in estimates of the number of metastable states (local minima of the Hamiltonian) in the theory of
spin glasses.  In fact, an integral
similar to that in Theorem   \ref{STRINGMAIN} arises in the formula for the expected number of local minima
of a random spin glass Hamiltonian.  Both heuristic and rigorous calculations lead to an exponential growth rate of the number of local minima
as the number of variables tends to infinity (see e.g. \cite{F} for
a mathematical discussion and references to the literature). The mathematical similarity of the problems
at least raises the question whether the number of string/M vacua should grow exponentially in the number $2 b_3$ of variables
$(G, Z)$, i.e. in the  `topological complexity' of the Calabi-Yau manifold $X$.

Our results do not settle this problem, and indeed it seems to be a difficult question. Here are some
of the difficulties: First,
  in  regard to the B\'ezout estimate, the naive
 argument ignores the fact that the critical point equation is a real $C^{\infty}$ equation, not a holomorphic
 one and so the B\'ezout estimate could be quite inaccurate. Moreover, a flux superpotential is not a polynomial
 and it is not  clear what `degree' it has, as measured by  its number critical points.
  In simple examples (see e.g.
 \cite{AD, DD, DGKT}, the superpotentials do not have many critical points and it is rather the large number of
 fluxes satisfying the tadpole constraint which produces the leading term $L^{b_3}.$ This is why
 the flux  $G$ has to be regarded as one of the variables if one wants to rescue the naive counting argument. In addition,
 the tadpole constraint has a complicated dimensional dependence. It
 induces a constraint on the inner integral in Theorem \ref{STRINGMAIN} to an ellipse in $b_3$ dimensions, and
 the volume of such a domain shrinks at the rate $1/(b_3)!$.  Further,  the volume of the Calabi-Yau moduli space
 is not known, and could be very small. Thus, there are a variety of competing influences on the growth rate of
 the number of vacua in $b_3$ which all have a factorial  dependence on the dimension.

To gain a better perspective on these issues,   it is important to
 estimate the integral giving the leading coefficient and  the remainder in Theorem \ref{STRINGMAIN}.
  The  inner  integral  is essentially an  integral of a homogeneous function of degree $b_3$ over an ellipsoid in
 $b_3$ dimensions, and is therefore very sensitive to the size of $b_3$. The full integral over moduli space carries
 the additional problem of estimating its volume.
Further, one needs to estimate how large $L$ is for a given $X$. Without such effective bounds
on $L$, it is not even possible to say whether any vacua exist which are consistent with known physical
quantities such as the cosmological constant.

\section{Sketch of proofs}

The proof of Theorem \ref{STRINGMAIN}  is in part an application of a lattice point result to the lattice of flux superpotentials. In addition,
it uses the formalism on the density of critical points of Gaussian random holomorphic sections in \cite{DSZ}.
The lattice point problem is to study the   distribution
of radial projections of lattice points in the shell $0 \leq Q[G] \leq L$  on the surface $Q[G] = 1.$
Radial projections arise because
the critical point equation $\nabla W_G = 0$ is
 homogeneous in $G$.

Thus, we consider the model problem:
 Let $\qcal \subset \R^n$ ($n\ge 2)$ be a   smooth,
star-shaped  set with $0\in \qcal^\circ$ and whose boundary has a non-degenerate second fundamental form.  Let $|X|_{\qcal}$ denote the
norm of $X\in \R^n$ defined by $\qcal =\{X\in\R^n:|X|_{\qcal}<1 \}\,.$ In the following, we denote the large parameter
by $\sqrt{L}$ to maintain consistency with Theorem \ref{STRINGMAIN}.

\begin{theo} \cite{DSZ3} \label{localvdC}  If $f$ is homogeneous of degree $0$ and $f|_{\d Q} \in C^{\infty}_0(\d Q)$, then  $$S_f(L) :
= \sum_{k \in \Z^n\cap \sqrt{L}\qcal\sm\{0\}} f(k) =L^{\frac{n}{2}} \int_{\qcal} f\,dX + O(L^{
\frac{n}{2} - \frac{n}{n+1}}), \;\; L \to \infty. $$
\end{theo}

   Although we have only stated it for smooth $f$,
the method
can be generalized  to $f |_{\d Q} = \chi_K$ where $K$ is a smooth domain in $\d Q$ \cite{Z2}. However, the remainder
then depends on $K$ and reflects the extent to which projections of lattice points  concentrate on $\d K \subset \d Q.$
The asymptotics are  reminiscent of the the result of van der Corput, Hlawka, Herz and Randol on the number
of lattice points in dilates of a convex set, but as of this time of writing we have not located
any prior studies of the radial projection  problem.
 Number theorists have
however studied the distribution of lattice points lying exactly on spheres (Linnik, Pommerenke). We also refer the interested reader to \cite{DO} for a
 recent article counting lattice points  in certain rational cones using methods of automorphic forms, in particular
 $L$-functions.
We thank B. Randol for some discussions of this problem; he  has informed the author that the result can also be extended to more general kinds of surfaces with degenerate second
fundamental forms.

Applying Theorem \ref{localvdC} to the string/M problem gives that
\begin{equation} \label{localvdCform} \ncal_{\psi}(L) = L^{b_3}\left[
 \int_{\{Q[W]\le 1\}}\langle C_W, \psi \rangle \,dW
   + O\left(L^{-\frac{2 b_3}{2b_3+1}}\right)\right]. \end{equation}
We then write   (\ref{localvdCform})  as an integral over the incidence relation (\ref{ANOTHERICAL})
 and change the order of integration to  obtain the leading coefficient
\begin{equation} \label{INTEGRAL} \int_{\{Q[W]\le 1\}}\langle C_W, \psi \rangle \,dW = \int_{\ccal}
\int_{\fcal_{z, \tau}} \psi(W, z, \tau) |\det H^c  W(z, \tau)|
  \chi_{Q_{z, \tau}} dW d V_{WP}(z, \tau)\end{equation}
  in Theorem \ref{STRINGMAIN}.  Heuristically,  the integral on the left side is given by
\begin{equation} \int_{\fcal} \int_{\ccal} \psi(W, Z) |\det H^c  W(Z)| \delta (\nabla W(z)) \chi_{Q}(Z) dW dV_{WP}(Z). \end{equation}
The factor  $|\det H^c W(Z)|$ arises in  the pullback of $\delta$ under $\nabla W(Z)$ for fixed $W$, since
it weights each term of (\ref{CN}) by $\frac{1}{|\det H^c  W(Z)|}$. We obtain the stated form of the integral in
(\ref{INTEGRAL}) by integrating first in $W$ and using the formula for the pull-back of a $\delta$ function under a linear
submersion.  That formula also  contains another factor $\frac{1}{\det A(Z)}$ where
 $A(Z) = \nabla_{Z'_j} \nabla_{Z''_k} \Pi_Z(Z', Z'')|_{Z' = Z'' = Z}$, where  $\Pi_Z$ is the \szego kernel of $\fcal_Z$, i.e.
 the orthogonal projection onto that subspace. Using special geometry, the matrix turns out to be just $I$ and hence the
 determinant is one.

 \section{\label{FURTHER} Other formulae for the critical point density}

In view of the difficulty of estimating the leading term in Theorem \ref{STRINGMAIN}, it is useful
to have alternative expressions. We now state two of them.

The first method is  to change variables to the Hessian $H^c
W(Z)$  under the Hessian map \begin{equation}
\label{HSUBZ} H_Z: \scal_{Z} \to \sym(m, \C) \oplus \C,\;\;\;
H_Z(W) = H^c W(Z), \end{equation}  where $m = \dim
\ccal=h^{2,1}+1$. It turns out that
Hessian map is an isomorphism to a real $b_3$-dimensional space
$\hcal_Z\oplus\C$, where
 \begin{eqnarray}\label{HCALZ} \hcal_Z = \mbox{span}_\R \left\{ \left(
 \begin{array}{cc}
0&   e_j    \\
     e_j^t  & \fcal^j(z)
 \end{array}
 \right), \ \left(
 \begin{array}{cc}
0&  i e_j    \\
     ie_j^t  &-i \fcal^j(z)
 \end{array}
 \right)\right \}_ {j = 1, \dots, h^{2,1}}\ .
\end{eqnarray}
Here, $e_j$ is the $j$-th standard basis element of $\C^{h^{2,1}}$
and
 $\fcal^j(z) \in \sym(h^{2,1}, \C)$ is the matrix $ \left( \fcal^j_{i
k} (z) \right)$ whose entries define the  so-called `Yukawa couplings'  (see
 \cite{St3, Can1} for the definition). We define the positive definite  operator $C_Z: \hcal_Z
 \to \hcal_Z$  by:
\begin{equation}\label{defC}(C_Z^{-1} H_Z W, H_Z W ) = Q_Z(W, \overline W).
\end{equation}
The entries in $C_Z$ are quadratic expressions in the $\fcal^j_{i k}$ (see \cite{DSZ3}).

\begin{theo} \label{PF} We have:
$$\begin{array}{lll}\kcal^\crit (Z)& = &\frac 1 {b_3!\det C_Z'} \int_{\hcal_Z \oplus \C}
\left|\det H^*H - |x|^2 I\right|\;\; e^{-(C\inv_Z H, H) +
|x|^2)}\,dH\,dx \;, \\ & & \\& = &\frac 1 {\det C_Z'}
\int_{\hcal_Z \oplus \C} \left|\det H^*H - |x|^2 I\right|
\chi_{C_Z} (H, x) dH dx,
\end{array} $$
where $\chi_{C_Z}$ is the characteristic function of the ellipsoid
$\{(C_Z H, H) + |x|^2) \leq 1\} \subset \hcal_Z.$
\end{theo}

Finally, we give  formula  of Itzykson-Zuber type as in
\cite[Lemma~3.1]{DSZ2},  which is useful in that it has  a fixed
domain of integration.

\begin{theo} \label{I} Let $\Lambda_Z = C_Z \oplus I$ on $\hcal_Z
\oplus \C$ and let $P_Z$ denote the orthogonal projection from
$\sym(m, \C)$ onto $\hcal_Z$. Then:
\begin{equation*}
\kcal^\crit(Z) = {c_m}\lim_{\ep' \to 0^+} \int_{\R^m } \lim_{\ep
\to 0^+}\int_{\R^m} \int_{\U(m)} \frac{\Delta(\xi)\,
\Delta(\lambda)\; |\prod_j \la_j| \,e^{i\langle \xi, \lambda
\rangle}   e^{- \epsilon |\xi|^2 -\epsilon'
|\lambda|^2}}{\sqrt{\det\left[i \La _Z P_Z \rho(g)^*\wh D(\xi) \rho(g)
+I\right]}}\, dg\, d \xi\, d\lambda,
\end{equation*}
 where:

\begin{itemize}

 \item
$m = h^{2,1} + 1, \;\;\;c_m=\frac{(-i)^{m(m-1)/2}}{2^m\,\pi^{2m}\,
\prod_{j=1}^mj!}\;$;

 \item $\Delta(\lambda) = \Pi_{i < j}
(\lambda_i - \lambda_j)$, \item $dg$ is unit mass Haar measure on
$\U(m)$, \item $\wh D(\xi)$ is the Hermitian operator on
$\sym(m,\C)\oplus \C$ given by
$$ \wh D(\xi) \big((H_{jk}),x\big) =
\left( \left(\frac{\xi_j+\xi_k}{2}\,H_{jk}\right),\
-\left(\textstyle\sum_{q=1}^m \xi_q\right) x \right)\;,$$ \item
 $\rho$ is the representation  of $\U(m)$ on
$\sym(m,\C)\oplus \C$ given by
$$\rho(g)(H, x) = (gHg^t,x)\;.$$
\item $\hcal_Z$ is a real (but not complex) subspace of $\sym(m, \C)$.
\end{itemize}

\end{theo}

The proof is similar to the one in \cite{DSZ2}, but we sketch the proof here to provide
a published reference. Some care must be taken since the Gaussian integrals are over real but not complex
spaces of complex symmetric matrices.

\begin{proof}

We first rewrite the integral in Theorem \ref{PF} as a  Gaussian
integral over $\hcal_Z \oplus \C$ (viewed as a real vector space):
$$\kcal^\crit (Z) =  \int_{\hcal_Z \oplus \C} |\det H^*H - |x|^2 I
|\;\;  \chi_{\{\langle \Lambda^{-1}_{Z}  H, H \rangle \leq 1\}} dH
dx  = \frac {\pi^{m} \sqrt{\det  \La_Z}}{b_3!}\; \ical(Z)\,, $$ where

\begin{equation}\label{ical}\ical(Z)=\frac 1
{\pi^{m} \sqrt{\det\Lambda_Z}} \int_{\hcal_{Z}\times \C}
\left|\det(HH^*-|x|^2I) \right| \exp\left( -{\langle \La_Z^{-1}(H,
x),(H, x) \rangle}\right) dH dx\;.\end{equation} Here, $H$ is a
complex $m \times m$ symmetric matrix, so $H^* = \overline{H}$. The inner product
in the exponent is the real part of the Hilbert-Schmidt inner product,
$\langle A, B \rangle = \Re \mbox{Tr} AB^*$.

 As in \cite{DSZ2}, we
rewrite the integral as
$$\ical(Z)= \lim_{\epsilon' \to 0}\; \lim_{\epsilon \to 0}
\ical_{\epsilon,\ep'} (Z)\;,$$ where $\ical_{\epsilon,\ep'}(Z)$ is
the absolutely convergent integral,
\begin{eqnarray}
\ical_{\epsilon,\ep'}(Z)  & = &
\frac{1}{(2\pi)^{m^2}\pi^{m}\sqrt{\det\La}} \int_{{\mathcal H}_m}
\int_{\hcal_m} \int_{\hcal_{Z}\times \C} \left|\det P\right| e^{-
\epsilon Tr \Xi^*
\Xi-\ep' Tr P^*P } e^{i \langle \Xi, P - HH^* +|x|^2I\rangle_{HS}}\nonumber\\
&&\quad\times\ \exp\left( -{\langle \La_Z^{-1}(H, x),(H, x)
\rangle}\right)\, dH \,dx \,dP \,d \Xi .\label{ical1a}
\end{eqnarray}

Here, $\hcal_m$ denotes the space of all Hermitian matrices of
rank $m$, and $\langle, \rangle_{HS}$ is the Hilbert-Schmidt inner product
$Tr A B^*$.   Formula \ref{ical1a} is valid, since as $\ep\to 0$, the
$d\Xi$ integral converges to the delta function $\de_{HH^*
-|x|^2I}(P)$. Then, as
 $\epsilon' \to 0$, the $d P$
integral evaluates the integrand at $P = H H^*-|x|^2I$ and we
retrieve the original integral $\ical(Z)$.

By the same manipulations as in \cite{DSZ2}, we obtain:
\begin{eqnarray}
\ical_{\epsilon,\ep'}(Z) & = &
\frac{(-i)^{m(m-1)/2}}{(2\pi)^m(\prod_{j=1}^mj!)\pi^{m}\sqrt{\det\Lambda_Z}}
\int_{\U(m)} \int_{\hcal_{Z}\times\C}
 \int_{\R^m} \int_{\R^m}\Delta(\lambda) \Delta(\xi)\,
 \left|\det(D(\lambda) )\right|\,
 \nonumber \\ &  \times & e^{i\langle \la,\xi\rangle}  e^{- \epsilon
(|\xi|^2 + |\lambda|^2)}
 e^{i \langle D(\xi),
 |x|^2 I  - g HH^* g^{*} \rangle_{HS} -{\langle
\La^{-1}(H, x),(H, x) \rangle}} \,d\xi\, d\la\, dH\, dx \, dg\;.
\label{ical3}
\end{eqnarray} Further we observe that the $dH dx$ integral
is a Gaussian integral.  Simplifying the phase as in \cite{DSZ2}
using
$$\langle D(\xi), gHH^*g^*-|x|^2 I\rangle_{HS}=
Tr (D(\xi) gHg^t \bar gH^*g^*) - Tr D(\xi)\,|x|^2 = \left\langle
\wh D(\xi)\rho(g)(H,x), \rho(g)(H,x)\right\rangle_{HS}$$ where $\wh
D(\xi)$ and $\rho(g)$ are as in the statement of the theorem, the
$\hcal_Z\times \C$ integral becomes \begin{equation} \label{NASTY}
\ical_{\xi,g}(Z):= \int_{\hcal_{Z} \times\C}
 \exp\big[-i \left\langle
\wh D(\xi)\rho(g)(H,x), \rho(g)(H,x)\right\rangle_{HS} -{\langle
\La_Z^{-1}(H, x),(H, x) \rangle}\big] \,dH\, dx. \end{equation}
The
 only new points in the calculation are that this Gaussian integral is over
 the Hessian space $\hcal_{Z}$ rather than over the full space
 of complex symmetric matrices of this rank, and that it is  a real subspace a complex vector space. Hence
 the Gaussian integral is a real one albeit with a complex quadratic form. We denote by
 $\pcal_Z$ the orthogonal projection
 $$\pcal_Z: \sym(m, \C) \to \hcal_Z $$
 and then we have:

\begin{eqnarray} \frac
{1}{\pi^m \sqrt{\det\La_Z}}\ical_{\xi,g}(Z)&=&\frac {1}{\sqrt{\det\La_Z}}\ \frac
{1}{\sqrt{\det[i \pcal_Z \rho(g)^*\wh D(\xi) \rho(g)
+\La _Z\inv]}} \nonumber\\
&=&   \frac {1}{\sqrt{\det\left[i \La _Z\pcal_Z \rho(g)^*\wh D(\xi)
\rho(g) +I_m\right]}}\;. \label{Hxint}\end{eqnarray} Substituting \eqref{Hxint} into
\eqref{ical3}, we obtain
 the desired formula.
 We now recall that $\Lambda = C' \oplus 1$. It follows that
$$\La _Z\pcal_Z \rho(g)^*\wh D(\xi)
\rho(g) +I_m = (C'_Z \pcal_Z \rho(g)^* D(\xi) \rho(g) +
I_{h^{21}}) \oplus (1 - \sum_{q = 1}^m \xi_q), $$ where
$$  D(\xi) (H_{jk})\big) =
\left(\frac{\xi_j+\xi_k}{2}\,H_{jk}\right).$$ Hence, its
determinant equals
$$(1 - \sum_{q = 1}^{h^{21}} \xi_q) \det (C'_Z \pcal_Z \rho(g)^* D(\xi) \rho(g) +
I_{h^{21}}). $$

 \end{proof}

\section{Black hole attractors}

We close this survey with a discussion of a simpler problem analogous to counting flux vacua that arises
in the quantum gravity of black holes \cite{St, FGK}, namely counting solutions of the black-hole attractor equation.
 For a mathematical introduction to this
equation, we refer the reader to  \cite{MM}. The attractor equation  is the same as the  critical point
equation for flux superpotentials except that  $\ccal = \mcal$ and  $G \in H^3(X, \R)$.
Physically, $\ncal_{\psi}(S)$ counts the so-called duality-inequivalent,
regular, spherically symmetric BPS black holes with entropy $S
\leq S_*.$ The charge of a black hole is an element $Q =
N^{\alpha} \Sigma_{\alpha} \in H^3(X, \Z)$. The central charge
${\mathcal Z} = \langle Q, \Omega \rangle$ plays the role of the
superpotential.

There are two main differences to the vacuum counting problems for flux superpotentials.
First, the reality of the flux $G$ in the black-hole attractor equation  $\nabla W_G(z) = 0 $ forces
$G\in H^{3,0}_z \oplus H^{0, 3}_z$ rather than $G \in H^{2,1}_z \oplus H^{0, 3}_z$ as in the flux vacua equation.
The space $ H^{3,0} \oplus H^{0, 3}$ is only $2$-dimensional and that drastically
simplifies the problem.
Second, by  a well-known computation due to Strominger, the
Hessian  $D \nabla G(z)$ of $|{\mathcal Z}|^2$ at a critical point is always a
scalar multiple $x \Theta$  of the curvature form of the line
bundle, which is the Weil-Petersson $(1,1)$ form.

We now state the analogue of Theorem \ref{PF} in the black hole attractor case (see also \cite{DD}).
The new feature is that the image of
 Hessian map from the space  $\scal_z$ of $W_G$ with a critical point at $z$ is the one-dimensional
 space of Hessians of the form
 \begin{equation}\begin{pmatrix} 0 &- x \Theta\\[8pt]
-\overline{x\Theta} & 0
\end{pmatrix}\;,
\end{equation}
 and hence the pushforward under the Hessian map  truly simplifies the integral in Theorem \ref{PF}.
  The  formula for the black-hole density becomes
$$\kcal^\crit_{\gamma, \nabla}(z)  =  \int_{ \C} |x|^{2 b_3} \chi_{Q_{z}}(x)
dx.
$$
We note that the difficult absolute value in Theorem \ref{PF} simplifies to a perfect square  in the black hole density
formula and can therefore be evaluated as a Gaussian integral. Additionally, the one-dimensionality of the space of
Hessians has removed the complexity of the  $b_3$-dimensional integral in the flux vacuum setting.

We can further simplify the integral by removing $Q_z$, which is a scalar multiple
of the Euclidean $|x|^2$.  The scalar multiple involves the orthogonal projection $\Pi_{\scal_z}(z, w)$ onto the space of
$\scal_z$ for the inner product $Q_z$.
If we change variables $x \to \sqrt{\Pi_{\scal_z}(z,z)}$, we get
$$\kcal^\crit_{\gamma, \nabla}(z) = |\Pi_{\scal_z}(z,z)|
 \int_{ \C} |x|^{2 b_3} e^{- \langle  x, x \rangle} dx.$$

  In \kahler  normal coordinates, use of special geometry
 shows that   $\Pi_z (z, z) = 1$. A
 simple calculation shows:
\begin{prop} The density of extremal black holes is given by:
$$\kcal^\crit_{\gamma, \nabla}(z)  = \frac{1}{ b_3} dV_{WP}
 \; \implies \; \; \ncal_{\psi}(L) \sim   L^{b_3} Vol_{WP}(\mcal). $$
\end{prop}

The analogy between the black hole
density and flux vacuum critical point density should be taken with some caution since the simplifying features
are likely to have over-simplified the problem.
We therefore mention another  modified flux vacuum problem in which the off-diagonal entries
$ x \Theta$ of the Hessian matrix vanish, so that the Hessian matrix is purely holomorphic and $\left|\det H^*H - |x|^2 \Theta\right|
= \left| \det H^*H \right|$
again becomes a perfect square which can be evaluated by the Wick method. Namely, if one uses a flat meromorphic connection
$\nabla$ rather than the Weil-Petersson connection, the curvature vanishes away from the polar divisor. The Weil-Petersson
connection arises naturally in string/M theory  \cite{CHSW}, but one may view a meromorphic connection as an approximation
in which the `Planck mass' is infinitely large. In any case,
it would be  interesting to evaluate the density of critical points relative
to meromorphic connections since they are more calculable and should have the same complexity as those for Weil-Petersson connections.

\end{document}